# 2-Variable Boolean Operation – Its Use in Pattern Formation

Sudhakar Sahoo

*Institute of Mathematics and Applications, Andharua, Bhubaneswar, 751003, INDIA*
Email: sudhakar.sahoo@gmail.com

Ipsita Mohanty, Garisha Chowdhary, Arpit Panigrahi

*Silicon Institute of Technology, Silicon Hills, Patia, Bhubaneswar*
Email: ipsita.mohanty689@gmail.com, garisha.pink@gmail.com, arpitpanigrahi@gmail.com

*Abstract* - **In this paper the theory of 2-Variable Boolean Operation (2-VBO) has been discussed on a pair of n-bit strings. 2-VBO serves to bring out the relation between numbers which when plot on a 2-D surface form interesting patterns; patterns that may be fixed, periodic, chaotic or complex. Some of these patterns represent natural fractals. This paper also provides mathematical analysis corresponding to each of the obtained patterns, which would aid to understanding their formation. 2-VBO is an attempt towards the production and classification of patterns which represent various mathematical models and naturally occurring phenomena.**

*Index Terms - 2-Variable Boolean Operation, Chaos and Fractals.*

## I. INTRODUCTION

A pattern is a design of natural or accidental origin that can be used in simulation of a phenomenon or problem, and serves as a model for predicting its future behavior. Once concepts like Cellular Automata, Fractals, L-Systems, Iterated Function Systems, etc. were introduced, it became clear that simple programs could produce behavior which is diverse and complex. Applying different rules one can obtain a variety of patterns. Although it is indeed true that for almost every rule the specific pattern produced is somewhat different, when one looks at all the rules together, one sees that the number of fundamentally different types of patterns produced is very limited.

In this paper we analyze the concept of 2-Variable Boolean Operation (2-VBO) based on the relation between two binary bits that can be used to transform numbers into patterns. Here, we have defined all sixteen 2-VBOs an extension work of Carry Value Transformation (CVT), Extreme Value Transformation (EVT) and Integer Value transformation (IVT) as defined in [4-7]. With these sixteen different mappings we have attempted to generate patterns which may be fixed, periodic, complex or chaotic. In our journey we have got various kinds of patterns, some of which can be classified as fractals.

The topics discussed in this paper are Wolfram's Classification of patterns [8, 9], discussed in Section II. The concept of 2-VBO is defined in section III. We have explored the formation of patterns using different 2-VBOs in section IV and a conclusion is made in section V.

## II. WOLFRAM'S CLASSIFICATION OF PATTERNS

The patterns formed in effect to all kinds of Cellular Automata rules can quite easily be assigned to one of just four basic classes, in order of increasing complexity in behavior of cells across states, as illustrated below.

*Class 1*: The behavior is simple, and almost all initial conditions lead to exactly the same uniform state.

*Class 2*: There are many different possible final states, but all of them consist of a certain set of simple structures that either remain the same forever or repeat every few steps.

*Class 3*: The behavior seems random in many respects, although triangles and other small-scale structures are essentially always at some level seen.

*Class 4*: It involves a mixture of order and randomness: localized structures are produced which on their own are fairly simple, but these structures move around and interact with each other in very complicated ways.

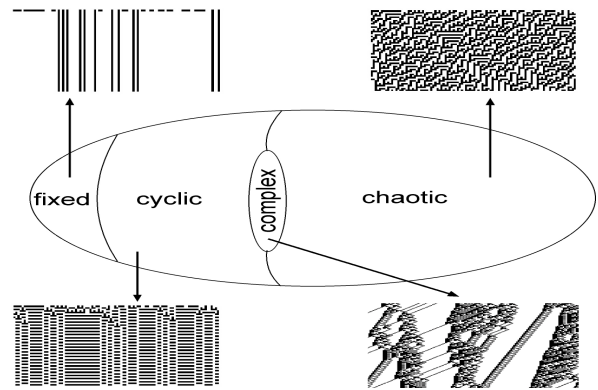

Figure 1. Shows Wolfram's Classification of patterns.

This classification by Wolfram was purely made on the basis of general visual appearance of the patterns. But later studies about the detailed properties of Cellular Automata revealed that these properties were closely correlated with the classes which had already been identified.

In general, class 1 and 2 systems rapidly settle down to states in which there are essentially no further activity. But class 3 systems continue to have many cells that change at every step, so that they in a sense maintain a high level of activity forever. Class 4 systems are then in the middle: for the activity that they show neither dies out completely, as in class 2, nor remains at the high level as seen in class 3.

### III. OPERATOR THAT RELATES PATTERNS WITH NUMBERS

Mathematics is the science with number as its language. In this section we have tried to make use of the beauty that lies in numbers to generate patterns, with 2-Variable Boolean Operator Table as the basis. In [4] Choudhury et al. gave a new operator namely Carry Value Transformation (CVT) which was able to generate Sierpinski triangle associating Boolean numbers with fractals. Here, we have defined sixteen 2-Variable Boolean Operations based on the relation between two binary bits. With these sixteen different mappings we have attempted to generate patterns, which may be fixed, periodic, complex or chaotic..

*A. 2-Variable Boolean Operation (2-VBO) and its Table (2-VBOT)*

The 2-Variable Boolean Operator Table (2-VBOT) is a set of 16 different Boolean functions defined on the set {0, 1}and 2-Variable Boolean Operation (2-VBO) operates upon the individual bits of two numbers (binary) depending upon the rule number (one of the 16 functions) given as input. Each rule is unique and the output of the operation varies in accordance with the various rules. Each function in 2-VBOT is a mapping from (B X B) → B, where B= {0, 1}. 2-VBO is performed upon corresponding bits of the binary equivalent of the two numbers. Let the bits be x and y. Since x and y can take values either 0 or 1, we can have $2^2$ i.e. four combinations for the string xy as (00, 01, 10, 11). Now for each of these four combinations we can have two bits, 0 or 1 as the resulting bit. So there are a total of ($2^4$ = 16) sixteen possible Boolean functions. Lets denote these functions as $f_0, f_1, f_2 \ldots f_{15}$. The functions are defined as $f_i$ = the binary equivalent of i for all i=0,…,15.

TABLE 1: . SHOWS 2-VARIABLE BOOLEAN OPERATOR TABLE 2-VBOT.

| x | y | $f_0$ | $f_1$ | $f_2$ | $f_3$ | $f_4$ | $f_5$ | $f_6$ | $f_7$ | $f_8$ | $f_9$ | $f_{10}$ | $f_{11}$ | $f_{12}$ | $f_{13}$ | $f_{14}$ | $f_{15}$ |
|---|---|---|---|---|---|---|---|---|---|---|---|---|---|---|---|---|---|
| 0 | 0 | 0 | 1 | 0 | 1 | 0 | 1 | 0 | 1 | 0 | 1 | 0 | 1 | 0 | 1 | 0 | 1 |
| 0 | 1 | 0 | 0 | 1 | 1 | 0 | 0 | 1 | 1 | 0 | 0 | 1 | 1 | 0 | 0 | 1 | 1 |
| 1 | 0 | 0 | 0 | 0 | 0 | 1 | 1 | 1 | 1 | 0 | 0 | 0 | 0 | 1 | 1 | 1 | 1 |
| 1 | 1 | 0 | 0 | 0 | 0 | 0 | 0 | 0 | 0 | 1 | 1 | 1 | 1 | 1 | 1 | 1 | 1 |

In Table 1 the function number uses Wolfram's naming convention [12], which is the decimal value of the output binary column vector from bottom to top.

To perform 2-VBO between two decimal numbers say p and q for a specified Boolean function $f_i$ (represented by rule i), following steps are to be followed:-

First, the binary equivalent of the two numbers is found out. Let these are $(p_1, p_2, p_3 \ldots p_m)$ and $(q_1, q_2, q_3 \ldots q_n)$ respectively. The length of these two binary equivalents is made equal by padding zeros to the left of the number, which is smaller in length. Then 2-VBO is done, applying the Boolean function $f_i$ from 2-VBOT, bitwise, for the given rule number i. The result so found is (say) $(b)_{10} = (b_1, b_2, b_3 \ldots b_k)_2$ where k = max (m, n).

Illustration:

Let us apply rule 2 (Boolean function $f_2$) to find 2-VBOT of the numbers $(43)_{10} \equiv (101011)_2$ and $(24)_{10} \equiv (11000)_2$. Since here $(101011)_2$ is having 6-bits and $(11000)_2$ has 5-bits, we have to pad one zero in the left side of $(11000)_2$. Then we perform the bit-wise operation to find 2-VBOT between them.

According to the Relation table, Rule 2 (Boolean function $f_2$) is defined as follows:

$f_2(0, 0) = 0, f_2(0, 1) = 1, f_2(1, 0) = 0, f_2(1, 1) = 0$

So the operation between the two numbers is as given below:

```
  1 0 1 0 1 1
  0 1 1 0 0 0
  ───────────
  0 1 0 0 0 0
```

So 2-VBOT of $(101011)_2$ and $(11000)_2 = (010000)_2 \equiv (16)_{10}$ and equivalently we can write 2-VBOT ( $(43)_{10}$, $(24)_{10}$ ) = $(16)_{10}$, when $f_2$ applied.

### IV. GENERATION OF PATTERNS USING 2-VBO

A matrix is constructed that contains only the values after applying 2-VBO as defined above, between all possible integers p and q arranged in an ascending order along x and y-axis respectively. We observe some interesting patterns formed by repetition of certain numbers in the matrix. Following are the steps that describe how matrix is constructed.

*Step 1*: A specific number of integers (say 0 to 15 as in table 2) are arranged in ascending order and placed to form rows (i) and columns (j) of a two dimensional matrix M.

*Step 2*: Each element of M is computed (for a specific rule, say 2) as: $M_{i,j}$ = 2-VBO ( i, j )

TABLE 2: PATTERN FORMATION, ASSIGNING RED TO 0 WHEN RULE 2 APPLIED

| | 0 | 1 | 2 | 3 | 4 | 5 | 6 | 7 | 8 | 9 | 10 | 11 | 12 | 13 | 14 | 15 |
|---|---|---|---|---|---|---|---|---|---|---|---|---|---|---|---|---|
| 15 | 15 | 14 | 13 | 12 | 11 | 10 | 9 | 8 | 7 | 6 | 5 | 4 | 3 | 2 | 1 | 0 |
| 14 | 14 | 14 | 12 | 12 | 10 | 10 | 8 | 8 | 6 | 6 | 4 | 4 | 2 | 2 | 0 | 0 |
| 13 | 13 | 12 | 13 | 12 | 9 | 8 | 9 | 8 | 5 | 4 | 5 | 4 | 1 | 0 | 1 | 0 |
| 12 | 12 | 12 | 12 | 12 | 8 | 8 | 8 | 8 | 4 | 4 | 4 | 4 | 0 | 0 | 0 | 0 |
| 11 | 11 | 10 | 9 | 8 | 11 | 10 | 9 | 8 | 3 | 2 | 1 | 0 | 3 | 2 | 1 | 0 |
| 10 | 10 | 10 | 8 | 8 | 10 | 10 | 8 | 8 | 2 | 2 | 0 | 0 | 2 | 2 | 0 | 0 |
| 9 | 9 | 8 | 9 | 8 | 9 | 8 | 9 | 8 | 1 | 0 | 1 | 0 | 1 | 0 | 1 | 0 |
| 8 | 8 | 8 | 8 | 8 | 8 | 8 | 8 | 8 | 0 | 0 | 0 | 0 | 0 | 0 | 0 | 0 |
| 7 | 7 | 6 | 5 | 4 | 3 | 2 | 1 | 0 | 7 | 6 | 5 | 4 | 3 | 2 | 1 | 0 |
| 6 | 6 | 6 | 4 | 4 | 2 | 2 | 0 | 0 | 6 | 6 | 4 | 4 | 2 | 2 | 0 | 0 |
| 5 | 5 | 4 | 5 | 4 | 1 | 0 | 1 | 0 | 5 | 4 | 5 | 4 | 1 | 0 | 1 | 0 |
| 4 | 4 | 4 | 4 | 4 | 0 | 0 | 0 | 0 | 4 | 4 | 4 | 4 | 0 | 0 | 0 | 0 |
| 3 | 3 | 2 | 1 | 0 | 3 | 2 | 1 | 0 | 3 | 2 | 1 | 0 | 3 | 2 | 1 | 0 |
| 2 | 2 | 2 | 0 | 0 | 2 | 2 | 0 | 0 | 2 | 2 | 0 | 0 | 2 | 2 | 0 | 0 |
| 1 | 1 | 0 | 1 | 0 | 1 | 0 | 1 | 0 | 1 | 0 | 1 | 0 | 1 | 0 | 1 | 0 |
| 0 | 0 | 0 | 0 | 0 | 0 | 0 | 0 | 0 | 0 | 0 | 0 | 0 | 0 | 0 | 0 | 0 |

*Step 3*: Then we look on the pattern of any integer(s) (0 in this case), and assign color (say red) to it (them), which form beautiful patterns (table 2). Out of them we observed some to be fractals.

. *In table 2, the pattern obtained using MATLAB 7.0 and after assigning color to matrix it has been rotated anticlockwise by∠90.*

Choosing different sequence of rows and columns in the matrix in step 1 or choosing different integer(s) to be colored, one can obtain uncountable number of patterns only by computing a single operation 2-VBO (a, b) in each entry position (a, b) of the table. We may also take the permutation of the integers taken along the axes and find various patterns.

In our attempt to find out patterns, we took the integers arranged in ascending order along both the rows and columns of a table. Then using step 2 we obtained different matrices for different rules. Observing the repetitive pattern of the numbers in the matrices, we assigned different colors to different numbers. Plotting the various matrices resulted in various patterns.

Some of the observed patterns are as follows:

**Note :** 1. Here each plot is a point 6 pixel thick and thus some overlapping or merging of adjacent positions might be seen when it is implemented in MATLAB 7.0.

2. *Rule: $i$ Red: $c_1$ Green: $c_2$ Blue: $c_3$ represents the function $f_i$ applied and the colors red, green and blue assigned to numbers $c_1$, $c_2$ and $c_3$ respectively*

A.   *Rule: 15 Red: 31 Green: 63 Blue: 127*

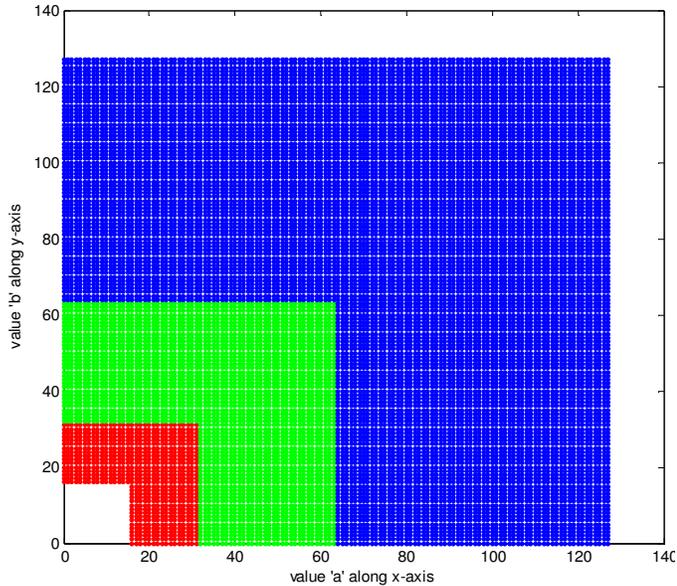

Figure 2.   Rule: 15 Red: 31 Green: 63 Blue: 127

The pattern in figure 2 is simple owing to the obvious reasons that rule 15 maps any binary combination to the bit '1' resulting in a number with all bits '1'. Thus in figure 3 any position (x, y), where $32 \leq x \leq 63$ and $32 \leq y \leq 63$ maps to $(111111)_2 = (63)_{10}$. Similarly any position in the range [64, 127] gives 127 and the range [16, 31] gives 31. Assigning different colors to all 1-bit, 2-bit, 3-bit, 4-bit numbers with all positions '1' will result in a similar pattern. This pattern with reference to Wolfram's classification is fixed in nature and thus belongs to Class 1.

B.   *Rule: 14 Red: 15 Green: 31 Blue: 63*

Referring 2-VBOT, rule 14 gives bit-0 for the binary combination 00 and bit-1 for the rest. It must be noted that it is the same as bitwise 'OR' operator.

In figure 3, the horizontal edge of the triangle in blue highlighted in the oval can be observed to be $y = 31$ for $x \in [32,63]$ which happens to be the resultant of the illustration below.

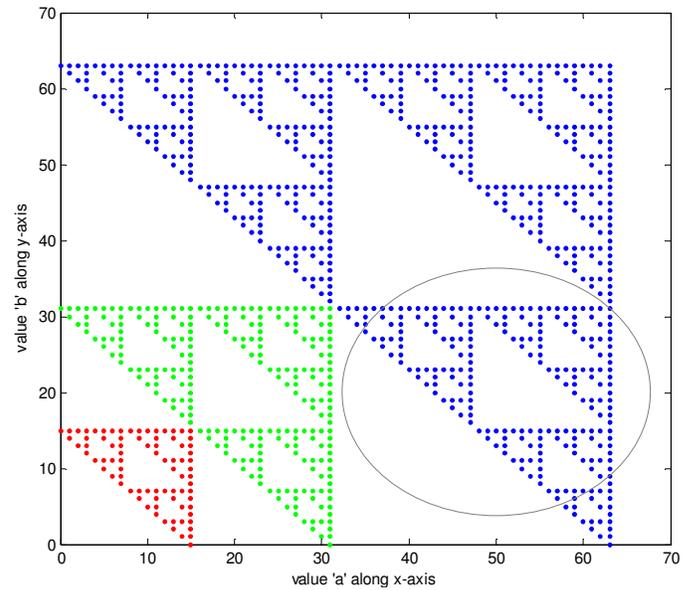

Figure 3.   Rule: 14 Red: 15 Green: 31 Blue: 63

| $a_5$ | $a_4$ | $a_3$ | $a_2$ | $a_1$ | $a_0$ | = a (x-axis) |
| $b_5$ | $b_4$ | $b_3$ | $b_2$ | $b_1$ | $b_0$ | = b (y-axis) |
|---|---|---|---|---|---|---|
| 1 | 1 | 1 | 1 | 1 | 1 | $\equiv (63)_{10}$ |

In order to get 63 as the 2-VBO value, if b = 31, then $a_5$ must be 1 (and the rest bits can take any value) i.e. a can take numbers in the range [32, 63].

Replacing a with b will give explanation for $x = 31$ and $y \in [32,63]$. For x = 63, we can have any number as the value for y (vertical edge of highlighted triangle). Similarly for y=63, any value for x is possible. 2-VBO(x, ~x) =63, where x is a 6 bit number and ~x is the bitwise compliment of x which gives the explanation for the third edge.

Similar explanations can be provided for 2-VBO(x, y) = 15(red), 2-VBO(x, y) = 31(green), as well for the edges of all other triangles. A keen observation shows that each of the figures in different colors represent the Sierpinski triangle (a deterministic exact self-similar fractal [1-3] belonging to Class 3) in different stages, with increase in number of bits (31→63→127 and so on).

C. *Rule: 13 Red: 31 Green: 63 Blue: 127*

Referring 2-VBOT, rule 13 gives bit-0 for the binary combination 01 and bit-1 for the rest.

*1)* The vertical edge in red i.e. $x = 31$ for $y \in [0,31], [32,63], [96,127]$

*Explanation*

$$\begin{array}{cc}
1\ 1\ 1\ 1\ 1 \equiv (31)_{10} & 0\ 1\ 1\ 1\ 1\ 1 \equiv (31)_{10} \\
\underline{b_4\ b_3\ b_2\ b_1\ b_0 = \ b} & \underline{1\ b_4\ b_3\ b_2\ b_1\ b_0 = b1} \\
1\ 1\ 1\ 1\ 1 \equiv (31)_{10} & 0\ 1\ 1\ 1\ 1\ 1 \equiv (31)_{10}
\end{array}$$

$$\begin{array}{c}
0\ 0\ 1\ 1\ 1\ 1\ 1 \equiv (31)_{10} \\
\underline{1\ 1\ b_4\ b_3\ b_2\ b_1 b_0 = b2} \\
0\ 0\ 1\ 1\ 1\ 1\ 1 \equiv (31)_{10}
\end{array}$$

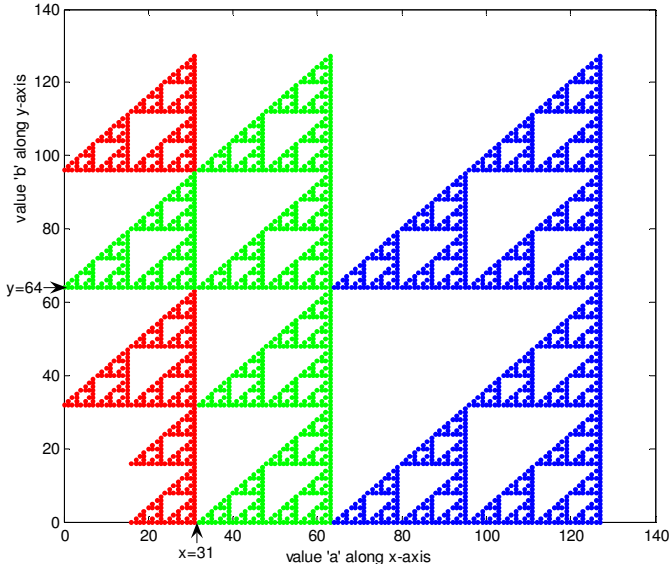

Figure 4. Rule: 13 Red: 31 Green: 63 Blue: 127

When x=31, in order to get 2-VBO(x, y) = 31, it is needed that $b \in [0,31], b1 \in [32,63]$ and $b2 \in [96,127]$ in the binary operations shown above.

*2)* Similarly, for y = 64 ≡ (1000000)$_2$, 2-VBO(x, y) = 63 ≡ (0111111)$_2$, it should be $x \in [0,63]$ illustrating the horizontal line in green and for y = 64 ≡ (1000000)$_2$, 2-VBO (x, y) =127≡ (1111111)$_2$, it should be $x \in [64,127]$ illustrating the horizontal line in blue. Alike illustrations explain other lines. The pattern so obtained is chaotic.

D. *Rule: 12 Red: 7 Green: 15 Blue: 31*

Explanation for the Vertical Bars plotted at x = 7, 15, 31 in figure 5:

From 2-VBO, $f_{12}(0, 0)=0$, $f_{12}(0, 1)=0$, $f_{12}(1, 0)=1$, and $f_{12}(1, 1)=1$ Here it can be seen that for $f_{12}$, 2-VBO(x, y) = x. So for some value of x=k and any value of y, the result of 2-VBO is k, resulting in a straight line x=k. The pattern so obtained is fixed.

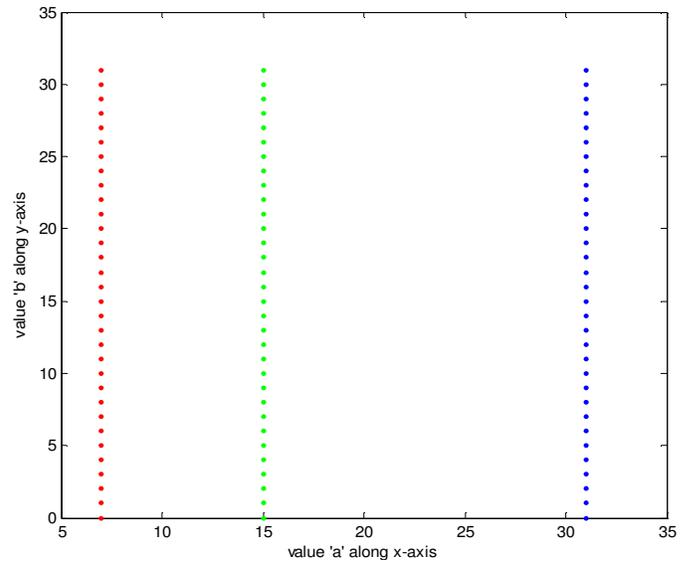

Figure 5. Rule: 12 Red: 7 Green: 15 Blue: 31

E. *Rule: 11 Red: 31 Green: 63 Blue: 127*

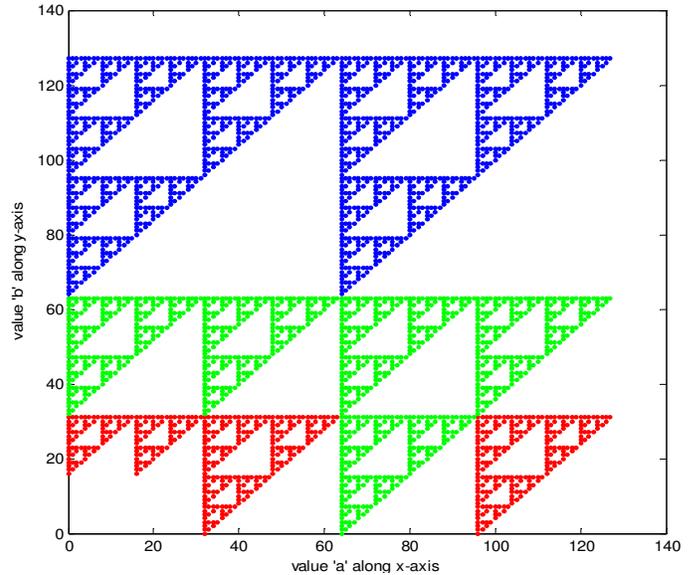

Figure 6. Rule: 11 Red: 31 Green: 63 Blue: 127

From 2-VBOT, $f_{13}(0,1) = 0$; on the contrary $f_{11}(1,0) = 0$ while they yield bit-1 for the rest of the bit combinations. Thus if x will be replaced by y, and vice versa in 2-VBO for rule 13, we will get figure 6, shown here for rule 11, which too belongs to the chaotic class.
.

F. *Rule: 10 Red: 7 Green: 15 Blue: 31*

Explanation for the Horizontal Bars plotted at y = 7, y = 15 and y = 31 in figure 7:

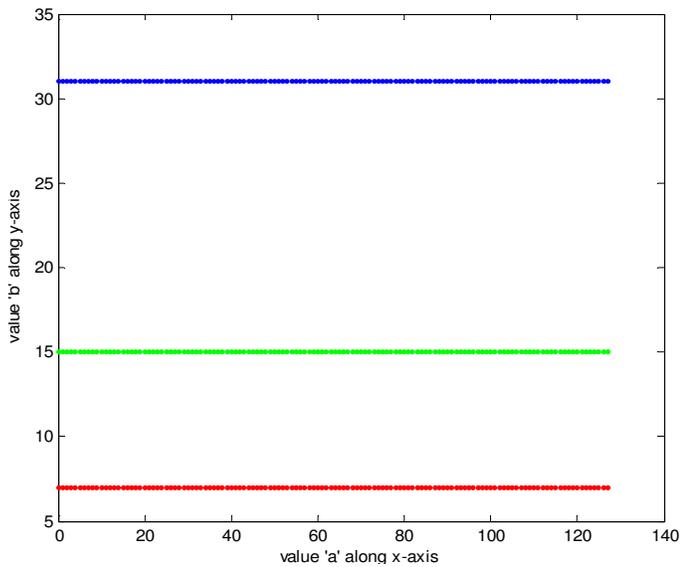

Figure 7. Rule: 10 Red: 7 Green: 15 Blue: 31

From 2-VBOT, $f_{10}(0, 0)=0$, $f_{10}(0, 1)=1$, $f_{10}(1, 0)=0$, and $f_{10}(1, 1)=1$. Here it can be seen that for $f_{10}$, 2-VBO(x, y) = y. So for some value, y=k and x taking any value, the result of 2-VBO is k, resulting in a straight line y=k. The pattern so obtained is fixed.

*G. Rule: 9 Red: 31 Green: 63 Blue: 127*

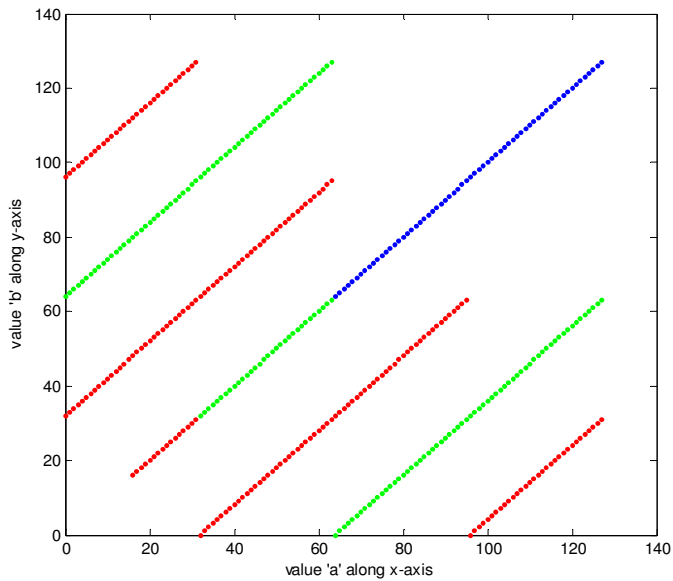

Figure 8. Rule: 9 Red: 31 Green: 63 Blue: 127

Explanation for 2-VBO(x, y) = 31(red) in the pattern shown in figure 8:

Referring 2-VBOT, rule 9 gives bit-1 for the binary combinations 00 and 11, and bit-0 for the rest. It must be noted that it is the same as bitwise 'XNOR' operator

*1)* For line x=y in red where $x \in [16, 31]$, since a and b take on the same binary bits, the resultant is a binary number with all bits 1, giving output $(31)_{10}$ as shown below:

$$\begin{array}{r} 1 \ a_3 \ a_2 \ a_1 \ a_0 = a \ (x\text{-axis}) \\ \underline{1 \ b_3 \ b_2 \ b_1 \ b_0} = b \ (y\text{-axis}) \\ 1 \ 1 \ 1 \ 1 \ 1 \equiv (31)_{10} \end{array}$$

*2)* For $x(\text{or } y) \in B_6$ and $y(\text{or } x) \in B_5$ gives a difference of $(32)_{10}$ between x and y resulting in the line x-y=32 (y-x=32) as the plot in red.

Similarly, the lines in blue and green can be illustrated. This pattern is periodic in nature i.e. belonging to Class 2.

*H. Rule: 8 Red: 0*

Referring 2-VBOT, rule 8 gives bit-1 for the binary combination 11, and bit-0 for the rest. It must be noted that it is the same as bitwise 'AND' operator.

It can be very clearly seen that as we scale up (16→32→64→128) in the figures 9 (a), 9 (b), 10 (a), 10 (b), we still get the same pattern and this will continue ad infinitum, which perfectly describes a deterministic fractal (discrete Sierpinski triangle as shown in figure 11). This can also be obtained using Rule 6 of 1-D Cellular Automata.

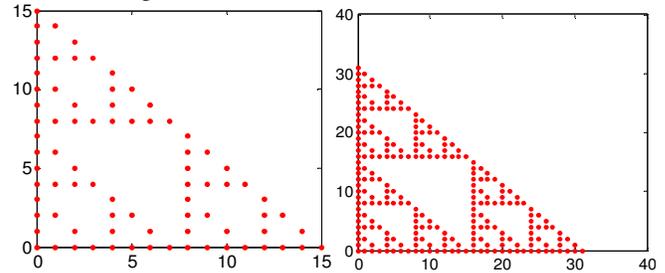

Figure 9. (a) Rule :8 Red: 0 scale: 32  (b) Rule :8 Red: 0 scale: 64

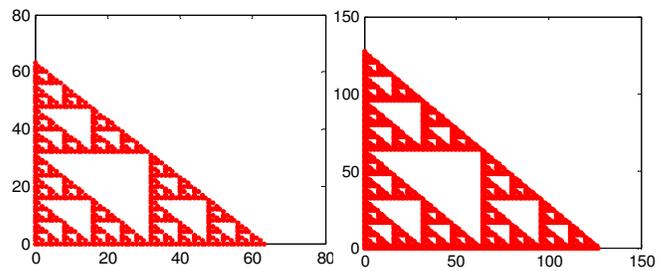

Figure 10. (a) Rule : 8 Red: 0 scale: 128  b) Rule :8 Red: 0 scale: 256

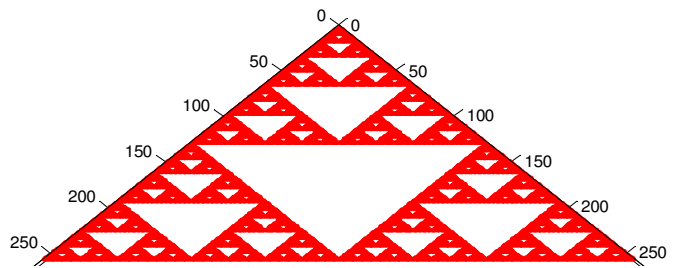

Figure 11. Figure 11 (b) rotated clockwise by ∠ 135

I. Rule: 7 Red: 31 Green: 63 Blue: 127

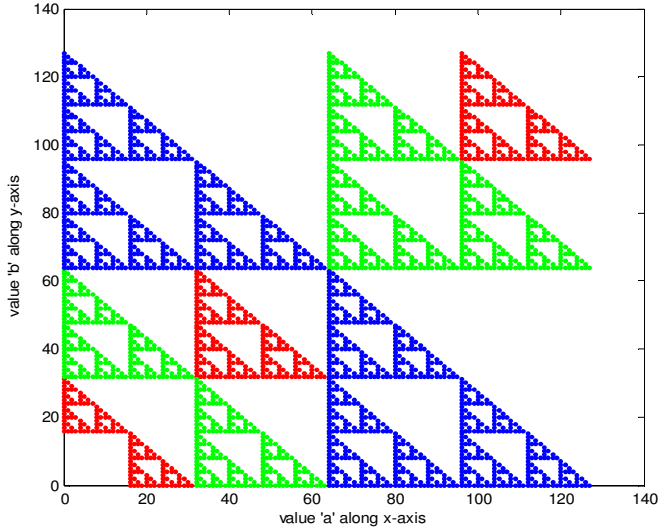

Figure 12. Rule: 7 Red: 31 Green: 63 Blue: 127

As per rule 7, $f_7(0, 0)=1$, $f_7(0, 1)=1$, $f_7(1, 0)=1$, and $f_7(1, 1)=0$, which is the NAND operation. It is the complement of rule-8. To get the red plot i.e. 2-VBOT(x, y) = 31 in figure 12, if x and y are 5-bit binary numbers, either (x<16 and y$\geq$16) or (x $\geq$ 16 and y<16), whereas for x and y as 6-bit or 7-bit numbers x must be equal to y. Similarly, the formation of green and blue triangles can be illustrated. The figure finally obtained belongs to Class 3 i.e. chaotic.

J. *Rule: 6 Red: 31 Green: 63 Blue: 127*

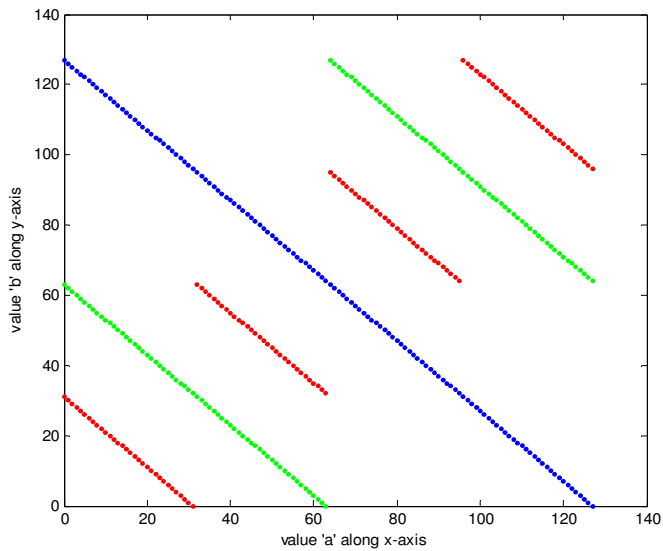

Figure 13. Rule: 6 Red: 31 Green: 63 Blue: 127

Referring 2-VBOT, rule 6 gives bit-1 for the binary combination 01 and 10, and bit-0 for the rest. It must be noted that it is the same as bitwise 'XOR' operator and thus complement of $f_9$. Hence the pattern obtained in figure 13 is similar and opposite in direction to that obtained in figure 8. This pattern is periodic in nature i.e. belonging to Class 2.

K. *Rule: 5 Red: 0 Green: 8 Blue: 16*

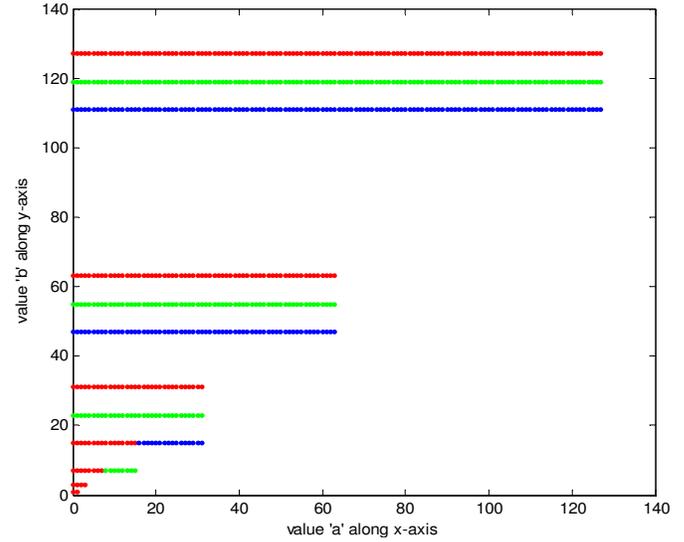

Figure 14. Rule: 5 Red: 0 Green: 8 Blue: 16

Explanation for the Horizontal Bars in figure 14:
$f_5(0, 0) =1$, $f_5(0, 1) =0$, $f_5(1, 0) =1$, $f_5(1, 1) =0$

Here it can be seen that for $f_5$, 2-VBO(x, y) = ~y (bit wise complement of y). To obtain 2-VBO(x, y) = 0, all bits of y must be 1. Hence the red bars are obtained at y=1, 3, 7, 15, 31, 63, 127... To obtain 2-VBO(x, y) = 8, all bits of y except the 4$^{th}$ bit must be 1. Hence the green bars are obtained at y=7, 23, 55, 119... Similarly, for 2-VBO(x, y) = 16, all bits of y except the 5$^{th}$ bit must be 1, resulting in blue bars at y=15, 47.111...This pattern is periodic with red bars occurring at an interval of $2^1, 2^2, 2^3, 2^4,....$ Similarly, green bars can be at interval of $2^4, 2^5,...$ and blue bars can be at interval of $2^5, 2^6,...$

L. *Rule: 4 Red: 0 Green: 32 Blue: 64*

Explanation for 2-VBO(x, y) = 64 (blue) in figure 15:
Referring to 2-VBOT, rule 4 gives bit 1 for the binary combination 10 and bit 0 for the remaining combinations.
*1)* The horizontal portion of the big blue triangle is observed to be y = 63 for x ranged from 64 to 128.
*Explanation:*

$$\begin{array}{ccccccc} a_6 & a_5 & a_4 & a_3 & a_2 & a_1 & a_0 \end{array} = a \text{ (x-axis)}$$
$$\underline{\begin{array}{ccccccc} 1 & 1 & 1 & 1 & 1 & 1 & 1 \end{array}} \equiv (63)_{10} \text{ (y-axis)}$$
$$\begin{array}{ccccccc} 1 & 0 & 0 & 0 & 0 & 0 & 0 \end{array} \equiv (64)_{10}$$

In order to get 32 as the 2-VBO value, if b=63, then $b_6$=0. So $a_6$ must be 1, while remaining digits of a can be either 0 or 1. So a can be any value in the range [64, 128].

*2)* Replacing a with b will give the explanation for x=63 for y in the range [0, 64]. Similar explanations can be given for all triangles.

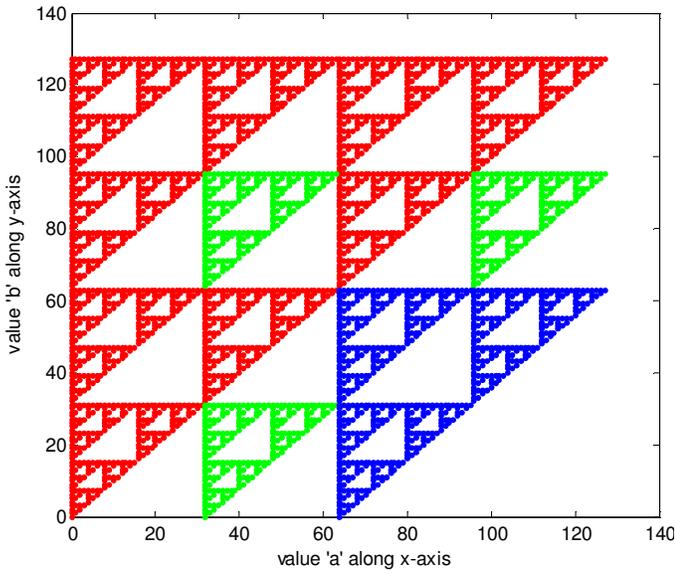

Figure 15. Rule: 4 Red: 0 Green: 32 Blue: 64

*M. Rule: 3 Red: 0 Green: 8 Blue: 16*

Explanation for the Vertical Bars plotted in figure 16:
$f_3(0, 0)=1$, $f_3(0, 1)=1$, $f_3(1, 0)=0$, $f_3(1, 1)=0$. Here it can be seen that for $f_3$, 2-VBO(x, y) = ~x. To obtain 2-VBO(x, y) = 0, all bits of x must be 1. Thus the red bars are obtained at x=1, 3, 7, 15, 31, 63, 127…To obtain 2-VBO(x, y) = 8, all bits of x except the 4$^{th}$ bit must be 1. Hence the green bars are obtained at x=7, 23, 55, 119…Similarly, for 2-VBO(x, y) = 16, all bits of x except the 5$^{th}$ bit must be 1, resulting in blue bars at x=15, 47, 111….This pattern is periodic with red bars occurring at an interval of $2^1$, $2^2$, $2^3$, $2^4$… Similarly, green bars can be at interval of $2^4$, $2^5$ … and green bars can be at interval of $2^5$, $2^6$…

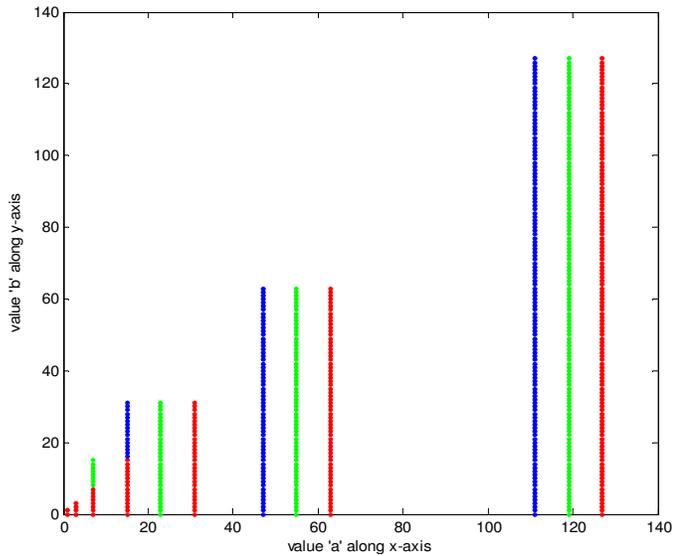

Figure 16. Rule: 3 Red: 0 Green: 8 Blue: 16

*N. Rule: 2 Red: 0 scale: 32*

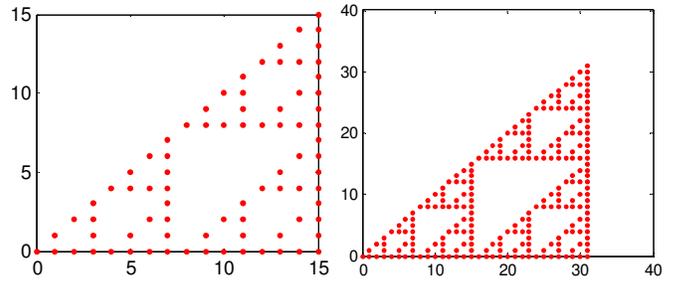

Figure 17. (a) Rule : 2 Red: 0 scale: 32 (b) Rule :2 Red: 0 scale: 64

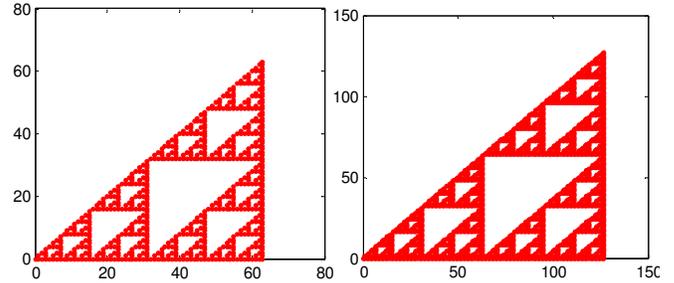

Figure 18. (a) Rule :2 Red: 0 scale: 128 (b) Rule :2 Red: 0 scale: 256

Referring 2-VBOT, rule 2 gives bit-1 for the binary combination 01, and bit-0 for the rest. Similar to that obtained in figure 9 and figure 10, as we scale up (16→32→64→128), we get the same pattern which will continue ad infinitum, perfectly describing a deterministic exact self similar fractal.

*O. Rule: 1 Red: 0*

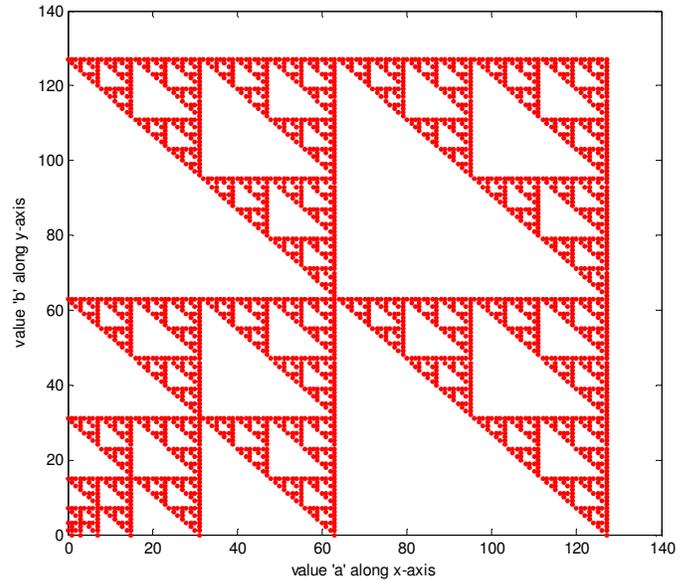

Figure 19. Rule : 1 Red: 0

Referring 2-VBOT, rule 1 gives bit-1 for the binary combination 00 and bit-0 for the rest i.e. rule-1 represents NOR operation. It can be clearly seen that rule 1 is the complement of rule 14.

In figure 3 (for rule 14), we had plot the numbers 15, 31 and 63. The complement of each of these numbers is zero. So plotting each of these numbers in rule 14 is equivalent to plotting zero in rule 1. Hence the figure obtained in this case is similar to that obtained for rule 14, thus belonging to the chaotic class.

*P.   Rule: 6 Red: 4, 6*

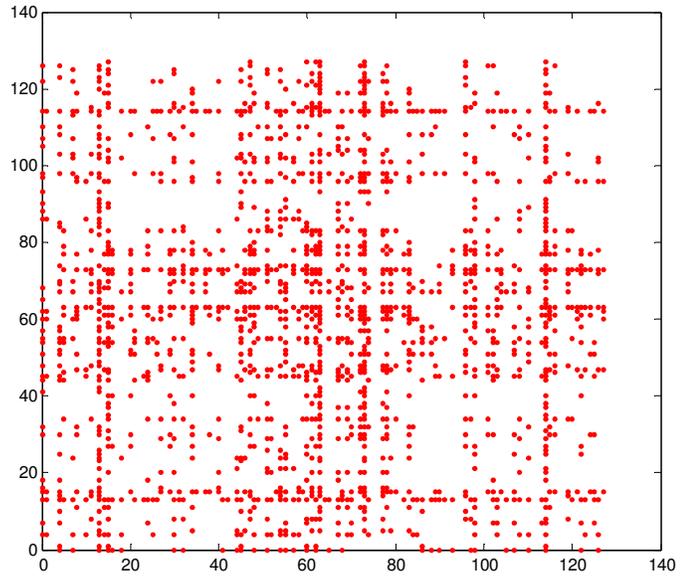

Figure 20.  Rule : 6 Red: 4, 6

Figure 20 was obtained after plotting a permutation of integers in the range [1, 128] along x-axis and y-axis and assigning color red to number 6 and 4. The pattern so obtained is complex i.e. belonging to Class 4.

It can be noted that on using integers in the range [1, 128] a total of 128! permutations are possible that can be plot along the axes. For each of these permutations we can assign to color to $^{128}C_0 + {}^{128}C_1 + {}^{128}C_2 + \ldots {}^{128}C_{128} = 2^{128}$ numbers and this can be done for all 16! rules, uniform and hybrid. Thus a total 16! x ($2^{128}$) x 128! number of patterns are possible for just the scale 128 and these patterns should be analyzed in the future.

## V.  CONCLUSION

This paper provides a thorough analysis of the 2-Variable Boolean Operations (2-VBO) which can form the basis to generate desired patterns applying different rules, uniform or hybrid, deciding the numbers along axes, and assigning different colors to numbers, as per our own requirement. It also furnished us with the fact that a simple operator performing operation on two binary bits can lead to generate patterns of all types, including chaotic and complex patterns that seem to be completely random. It was also successful in generating exact self-similar fractal, Sierpinski triangle in particular (in discrete form) through functions, $f_8$ and $f_2$ which can also be generated through Rule 6 in 1-D CA. We wish to extend our work, making exhaustive use of two numbers to generate all possible patterns and establish that 2-VBO can be used to simulate real life complex models like DNA sequence, stock market fluctuations etc. as well.

## ACKNOWLEDGMENT

The authors would like to thank our other co-researchers Prof. Pabitra Pal Choudhury, Sk. Sarif Hassan and Prof. Birendra Kumar Nayak for providing us the inspiration to extend further the work on Carry Value Transformation (CVT), Extreme Value Transformation (EVT) and Integer Value Transformation (IVT).